\documentclass[twocolumn,showpacs,preprintnumbers,amsmath,amssymb]{revtex4}
\usepackage{graphicx}% Include figure files
\usepackage{dcolumn}% Align table columns on decimal point
\usepackage{bm}% bold math

\begin{document}

\preprint{}
\title{Hybrid-order Poincar\'{e} sphere}% Force line breaks with \\
\author{Xunong Yi$^{1,2,3}$}
\author{Yachao Liu$^{1}$}
\author{Xiaohui Ling$^{2}$}
\author{Xinxing Zhou$^1$}
\author{Yougang Ke$^1$}
\author{Hailu Luo$^{1}$}\email{hailuluo@hnu.edu.cn}
\author{Shuangchun Wen$^{1}$}
\author{Dianyuan Fan$^{2}$}
\affiliation{$^1$ Laboratory for Spin Photonics, School of Physics
and Electronics, Hunan University, Changsha 410082, China}
\affiliation{$^2$ College of Optoelectronic Engineering, Shenzhen
University, Shenzhen 518060, China} \affiliation{$^3$ School of
Physics and Electronic Information Engineering, Hubei Engineering
University, Xiaogan 432000, China}

\date{\today}% It is always \today, today,
             %  but any date may be explicitly specified

\begin{abstract}
In this work, we develop a hybrid-order Poincar\'{e} sphere to
describe the evolution of polarization states of wave propagation in
inhomogeneous anisotropic media. We extend the orbital Poincar\'{e}
sphere and high-order Poincar\'{e} sphere to a more general form.
Polarization evolution in inhomogeneous anisotropic media with
special geometry can be conveniently described by state evolution
along the longitude line on the hybrid-order Poincar\'{e} sphere.
Similar to that in previously proposed Poincar\'{e} spheres, the
Berry curvature can be regarded as an effective magnetic field with
monopole centered at the origin of sphere and Berry connection can
be interpreted as the vector potential. Both the Berry curvature and
the Pancharatnam-Berry phase on the hybrid-order Poincar\'{e} sphere
are demonstrated to be proportional to the variation of total
angular momentum. Our scheme provides a convenient method to
describe the spin-orbit interaction in inhomogeneous anisotropic
media.

\end{abstract}

% PACS, the Physics and Astronomy
                             % Classification Scheme.

%Use showkeys class option if keyword
                              %display desired

\maketitle

Polarization and phase are two intrinsic features of electromagnetic
waves~\cite{Born1997}. Fundamental polarization states, such as
linear, circular, and elliptical polarizations, have a spatial
homogeneous distribution. In 1892, a prominent geometric
representation of polarization known as the Poincar\'{e} sphere is
proposed to describe the polarization state of light as a point on
the surface of a unit sphere~\cite{Poincare1892}. The Poincar\'{e}
sphere unifies the fundamental polarizations, where the polarization
states represented by a complex Jones vectors are mapped to the
sphere's surface through the Stokes parameters in the sphere's
Cartesian coordinates. This geometric characterization not only
greatly simplifies the calculations of geometric phase, but also
provides a deeper insight into physical mechanisms. As a result, the
representation of Poincar\'{e} sphere has become an important
technique to deal with the polarization evolution in different
physical systems.

Recently, the orbital Poincar\'{e} sphere has been proposed as a
geometrical construction to represent the state evolution in phase
space~\cite{Padgett1999}. In analogy to the space of polarization,
the north and south poles of the orbital Poincar\'{e} sphere
correspond to the Laguerre-Gaussian modes with opposite topological
charges. The points in the equator of the sphere correspond to
Hermite-Gaussian modes~\cite{Galvez2003,Karimi2010}. In past several
years, high-order solutions with a spatial inhomogeneous
polarization and phase have drawn much attention~\cite{Zhan2009}.
More recently, the high-order Poincar\'{e} sphere has been proposed
to describe the evolution of both polarization and
phase~\cite{Milione2011,Holleczek2011}. The north and south poles of
the high-order Poincar\'{e} sphere represent the opposite spin
states and orbital states. Any state on the high-order Poincar\'{e}
sphere, can be realized by a superposition of the two orthogonal
states~\cite{Milione2012,Cardano2012,Chen2014}. However, the
high-order solutions on high-order Poincar\'{e} sphere are still
confined to some special cases. Hence, it is necessary for us to
extend the orbital Poincar\'{e} sphere and high-order Poincar\'{e}
sphere to a more general form.

In this work, we develop a hybrid-order Poincar\'{e} sphere to
describe the evolution of phase and polarization of wave propagating
in inhomogeneous anisotropic media. We find that the representation
of the polarization states in inhomogeneous anisotropic media has a
similar expression of polarization states on high-order Poincar\'{e}
sphere~\cite{Milione2011,Holleczek2011} with only the orbital states
being different. This interesting property motivates us to develop a
hybrid-order Poincar\'{e} sphere to describe the evolution of
polarization states, and therefore extending the high-order
Poincar\'{e} sphere and orbital Poincar\'{e} sphere to a more
general form. It is known that the orbital states in the two poles
of orbital and high-order Poincar\'{e} spheres have the same value
but opposite signs. Unlike the previously reported cases, the
orbital states on hybrid-order Poincar\'{e} sphere should not be
confined to this certain condition and can be chosen arbitrarily. We
show that the polarization evolution in inhomogeneous anisotropic
media with special geometry can be conveniently described by state
evolution along the longitude line on the hybrid-order Poincar\'{e}
sphere. Furthermore, Berry connection, Berry curvature, and
Pancharatnam-Berry phase associated with the evolution of
polarization state are discussed.

\section{Hybrid-order Poincar\'{e} sphere}
We now develop a hybrid-order Poincar\'{e} sphere to describe the
evolution of polarization and phase in inhomogeneous anisotropic
media. It is assumed that the media are composed of local waveplates
whose optical axis directions are specified by a space-variant angle
\begin{equation}
\alpha(r,\varphi)=q \varphi +\alpha_0,
\end{equation}
where $r$ is the radial coordinate, $\varphi$ is the azimuthal
coordinate, $\alpha_0$ is a constant angle specifying the initial
orientation on the axis $x$, and $q$ is a constant specifying the
topological charge. The inhomogeneous birefringent elements having
specified geometry can be designated as
$q$-plates~\cite{Marrucci2006}.

Let us consider that the $q$-plate is illuminated by a circularly
polarized vortex wave
$|\mathbf{\psi}\rangle={\sqrt2}/2(\mathbf{\hat{e}}_x+i\sigma\mathbf{\hat{e}}_y)
\exp(il\varphi)$ with spin angular momentum (SAM)
$\sigma\hbar$~\cite{Beth1936} and orbit angular momentum (OAM)
$l\hbar$~\cite{Allen1992}, where $\sigma=+1$ for the left-handed
circular (LHC) polarization and $\sigma=-1$ for the right-handed
circular (RHC) one. The evolution of optical field in the
inhomogeneous medium can be obtained in Appendix~\ref{AppA} as
\begin{eqnarray}
|\mathbf{\psi}_{l,m}\rangle&=&\cos\frac{\delta}{2}
\frac{\sqrt2}{2}(\mathbf{\hat{e}}_x+i\sigma\mathbf{\hat{e}}_y)\exp(il\varphi)\nonumber\\
&&+\sin\frac{\delta}{2}\frac{\sqrt2}{2}(\mathbf{\hat{e}}_x-i\sigma\mathbf{\hat{e}}_y)\exp(im\varphi)\nonumber\\
&&\times\exp[i(2\sigma\alpha_0-\frac{\pi}{2})]\label{CVV}.
\end{eqnarray}
Here, $m=l+2\sigma{q}$. It should be noted that diffraction inside
the  $q$-plate is neglected, so that only evolution of polarization
and phase is taking place. This is valid as long as the thickness of
the medium is small as compared with a Rayleigh diffraction length.
The field in the $q$-plate can be regarded as a superposition of a
first wave that has the same SAM and OAM as the input one, and a
second wave having reversed SAM and a modified OAM given by
$m\hbar$. It means that the input wave only partially occurs
spin-to-orbital angular momentum conversion~\cite{Marrucci2008}. The
field amplitudes of the two components of the field depend on the
birefringent retardation $\delta$, and are given by $\cos\delta/2$
and $\sin\delta/2$, respectively. In addition, a singularity should
be generated around central region of $q$-plate.

\begin{figure}
\centerline{\includegraphics[width=8cm]{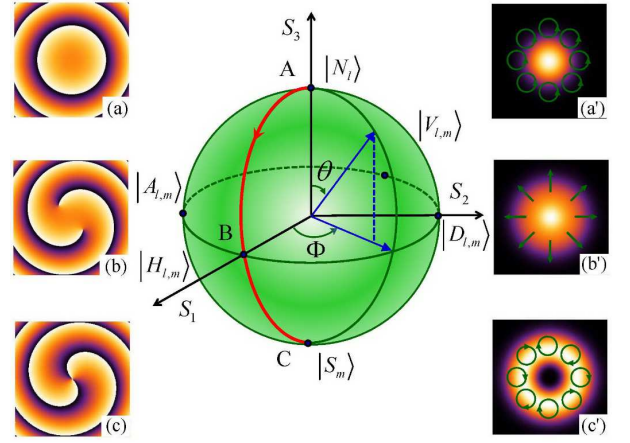}}
\caption{\label{Fig1} (Color online) Schematic illustration of the
evolution of phase and polarization on hybrid-order Poincar\'{e}
sphere. Insets (a)-(c) show the the phase for points A, B, and C,
respectively. Insets (a$'$)-(c$'$) show the polarization state of
the three points. Here, we assume the north pole with state
$\sigma=+1$ and $l=0$, while the south pole with $\sigma=-1$ and
$m=+2$.}
\end{figure}

We note that the field represented by Eq.~(\ref{CVV}) has a similar
expression to the field represented by high-order Poincar\'{e}
sphere~\cite{Milione2011,Holleczek2011} with only the orbital states
being different. It inspires us to develop a hybrid-order
Poincar\'{e} sphere to describe the evolution of phase and
polarization. The field for a monochromatic paraxial light beam can
be expressed as a two dimensional Jones vector:
\begin{equation}
|\mathbf{\psi}_{l,m}\rangle=\mathbf{\psi}_{N}^l|\mathbf{N}_l\rangle+\mathbf{\psi}_{S}^m|\mathbf{S}_m\rangle
\label{JonesV},
\end{equation}
where
\begin{equation}
|\mathbf{N}_l\rangle=\frac{\sqrt2}{2}(\mathbf{\hat{e}}_x+i\sigma\mathbf{\hat{e}}_y)\exp(il\varphi)\label{NLI},
\end{equation}
\begin{equation}
|\mathbf{S}_m\rangle=\frac{\sqrt2}{2}(\mathbf{\hat{e}}_x-i\sigma\mathbf{\hat{e}}_y)\exp(im\varphi)\label{SMI}.
\end{equation}
Here, $|\mathbf{N}_l\rangle$ and $|\mathbf{S}_m\rangle$ with
different topological charges construct an orthogonal polarization
basis.  Any polarization state on hybrid-order Poincar\'{e} can be
described as a superposition of the orthogonal bases with
coefficients $\mathbf{\psi}_{N}^l$ and $\mathbf{\psi}_{S}^m$,
respectively.

We now map the polarization states on hybrid-order Poincar\'{e}
sphere by representing the Stokes parameters in the sphere's
Cartesian coordinates. According to Eqs.~(\ref{NLI}) and
(\ref{SMI}), we redefine the Stokes parameters as~\cite{Born1997}
\begin{equation}
S_0^{l,m}=|\mathbf{\psi}_N^{l}|^2+|\mathbf{\psi}_S^{m}|^2
\label{S0},
\end{equation}
\begin{equation}
S_1^{l,m}=2|\mathbf{\psi}_N^{l}||\mathbf{\psi}_S^{m}|\cos\Phi
\label{S1},
\end{equation}
\begin{equation}
S_2^{l,m}=2|\mathbf{\psi}_N^{l}||\mathbf{\psi}_S^{m}|\sin\Phi
\label{S2},
\end{equation}
\begin{equation}
S_3^{l,m}=|\mathbf{\psi}_N^{l}|^2-|\mathbf{\psi}_S^{m}|^2
\label{S3},
\end{equation}
where $\Phi=\arg(\mathbf{\psi}_N^{l})-\arg(\mathbf{\psi}_S^{m})$,
$|\mathbf{\psi}_N^{l}|^2$ and  $|\mathbf{\psi}_S^{m}|^2$ are the
intensities of $|\mathbf{N}_l\rangle$ and $|\mathbf{S}_m\rangle$,
respectively. Using $S_1^{l,m}$, $S_2^{l,m}$, and $S_3^{l,m}$ as the
sphere's Cartesian coordinates, we construct a new Poincar\'{e}
sphere with $S_0^{l,m}$ the unit radius. Equations (\ref{NLI})  and
(\ref{SMI}) denote the states on two poles with orthogonal circular
polarizations. It is worth noting that $|\mathbf{N}_l\rangle$ and
$|\mathbf{S}_m\rangle$ generally have different topological charges,
i.e., $l\neq m$. Therefore, we term the new Poincar\'{e} sphere as
hybrid-order Poincar\'{e} sphere.

Generally, the equatorial points on the hybrid-order Poincar\'{e}
sphere represent a superposition of equal intensities of the two
orthogonal states.  The horizontal and vertical polarization basis
($|\mathbf{H}_{l,m}\rangle$, $|\mathbf{V}_{l,m}\rangle$) can be
obtained through the relations
$|\mathbf{H}_{l,m}\rangle=(|\mathbf{N}_l\rangle+|\mathbf{S}_m\rangle)/2$
and
$|\mathbf{V}_{l,m}\rangle=-i(|\mathbf{N}_l\rangle-|\mathbf{S}_m\rangle)/2$,
then we have
\begin{eqnarray}
|\mathbf{H}_{l,m}\rangle&=&\exp\frac{i(l+m)\varphi}{2}
\bigg[\cos\frac{(l-m)\varphi}{2}\mathbf{\hat{e}}_x\nonumber\\
&&+\sin\frac{(l-m)\varphi}{2}\mathbf{\hat{e}}_y\bigg]
\label{HOPSVLM},
\end{eqnarray}
\begin{eqnarray}
|\mathbf{V}_{l,m}\rangle&=&\exp\frac{i(l+m)\varphi}{2}\bigg[
\cos\left(\frac{l-m}{2}\varphi+\frac{\pi}{2}\right)\mathbf{\hat{\hat{e}}}_x\nonumber\\
&&+\sin\left(\frac{l-m}{2}\varphi+\frac{\pi}{2}\right)\mathbf{\hat{\hat{e}}}_y\bigg]
\label{VOPSDLM},
\end{eqnarray}
with coefficients
$\mathbf{\psi}_{H}^{l,m}=(\mathbf{\psi}_{N}^l+\mathbf{\psi}_{S}^m)/\sqrt{2}$
and
$\mathbf{\psi}_{V}^{l,m}=i(\mathbf{\psi}_{N}^l-\mathbf{\psi}_{S}^m)/\sqrt{2}$.
Similarly, the diagonal and antidiagonal polarization basis
$|\mathbf{D}_{l,m}\rangle$ and $|\mathbf{A}_{l,m}\rangle$ can be
obtained through the relations
$|\mathbf{D}_{l,m}\rangle=(|\mathbf{H}_{l,m}\rangle+|\mathbf{V}_{l,m}\rangle)/\sqrt{2}$
and
$|\mathbf{A}_{l,m}\rangle=(|\mathbf{H}_{l,m}\rangle-|\mathbf{V}_{l,m}\rangle)/\sqrt{2}$,
respectively, and we have
\begin{eqnarray}
|\mathbf{D}_{l,m}\rangle&=&\exp\frac{i(l+m)\varphi}{2}\bigg[
\cos\left(\frac{l-m}{2}\varphi+\frac{\pi}{4}\right)\mathbf{\hat{e}}_x\nonumber\\
&&+\sin\left(\frac{l-m}{2}\varphi+\frac{\pi}{4}\right)\mathbf{\hat{e}}_y\bigg]
\label{HOPSDLM},
\end{eqnarray}
\begin{eqnarray}
|\mathbf{A}_{l,m}\rangle&=&\exp\frac{i(l+m)\varphi}{2}\bigg[
\cos\left(\frac{l-m}{2}\varphi+\frac{3\pi}{4}\right)\mathbf{\hat{e}}_x\nonumber\\
&&+\sin\left(\frac{l-m}{2}\varphi+\frac{3\pi}{4}\right)\mathbf{\hat{e}}_y\bigg]
\label{HOPSALM},
\end{eqnarray}
with coefficients
$\mathbf{\psi}_{D}^{l,m}=(\mathbf{\psi}_{H}^{l,m}+\mathbf{\psi}_{V}^{l,m})/\sqrt{2}$
and
$\mathbf{\psi}_{A}^{l,m}=(\mathbf{\psi}_{H}^{l,m}-\mathbf{\psi}_{V}^{l,m})/\sqrt{2}$.
Note that the equatorial points represent vector vortex waves and
the relative phase of the superposition determines the orientation
of the longitude on equator.

\begin{figure}
\centerline{\includegraphics[width=8cm]{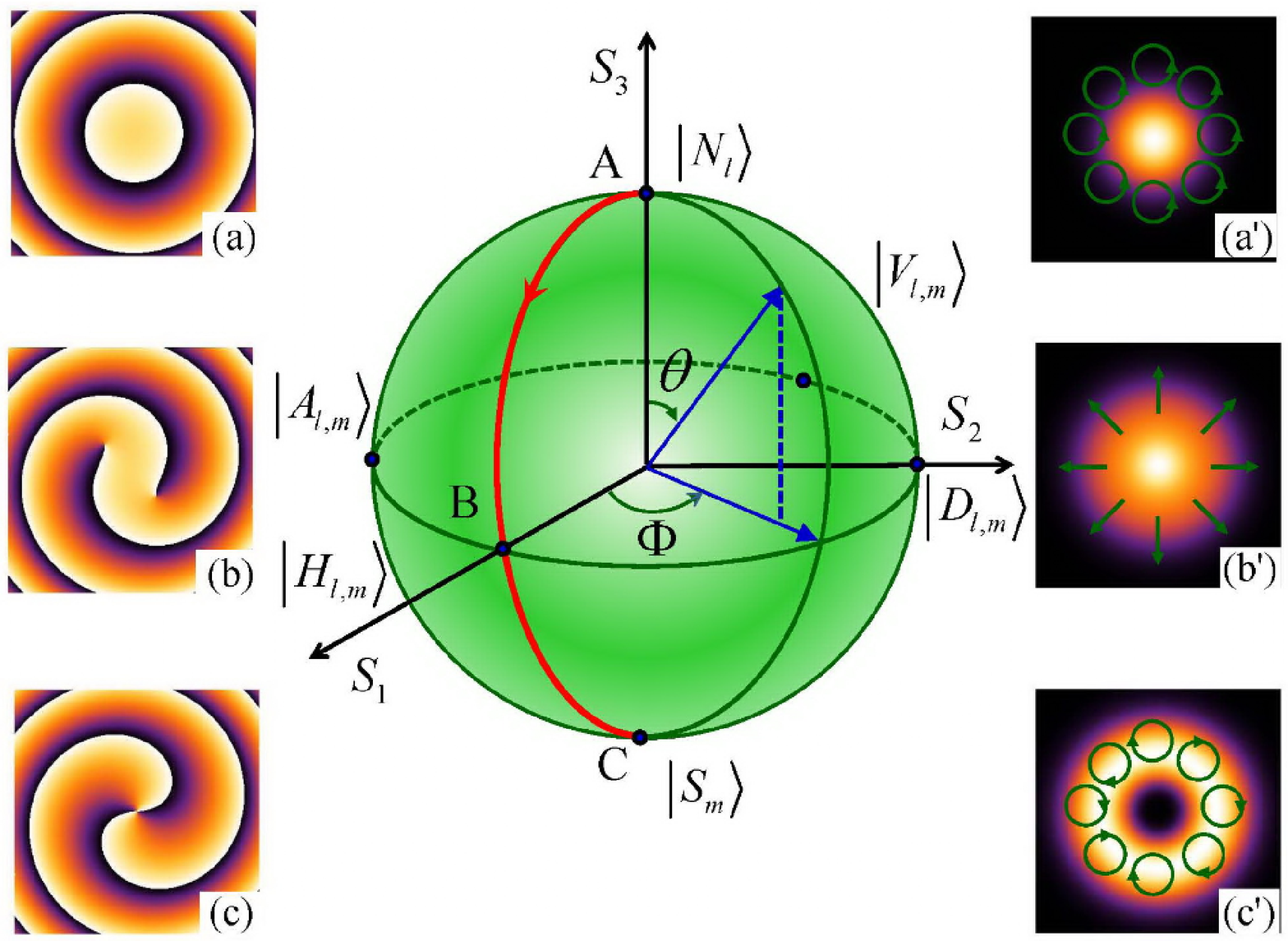}}
\caption{\label{Fig2} (Color online)  Schematic illustration of the
evolution of phase and polarization on hybrid-order Poincar\'{e}
sphere. Insets (a)-(c) show the evolution the phase for points $A$,
$B$, and $C$, respectively. Insets (a$'$)-(c$'$) show the evolution
of polarization of the three points. Here, we assume the noth pole
with state $\sigma=+1$ and $l=0$, while the south pole with
$\sigma=-1$ and $m=-2$.}
\end{figure}

Figures~\ref{Fig1} and~\ref{Fig2} show the two cases for the
evolution of phase and polarization on hybrid-order Poincar\'{e}
spheres. Comparing with the high-order Poincar\'{e} sphere, the
hybrid-order Poincar\'{e} sphere has two special features: (1) The
orbital states on hybrid-order Poincar\'{e} sphere should not be
confined to have the same value and opposite signs. As a result, the
cylindrical vector vortex beam can be represented by the equatorial
points. Intermediate points between the poles and equator represent
elliptically polarized vector vortex beam. (2) Polarization and
phase evolution in any $q$-plate can be conveniently described by
state evolution along the longitude line on the hybrid-order
Poincar\'{e} sphere. These features will be described in detail in
the next section.

\section{Berry connection, Berry curvature, and geometric phase}
From a basic geometric transformation on the Poincar\'{e} sphere,
the factor of $\varphi\rightarrow\Phi/2$ is a consequence of
transformation between the physical SU(2) space of the light beam
and the topological SO(3) space of the hybrid-order Poincar\'{e}
sphere~\cite{Milione2012}. For a monochromatic wave, the
polarization states can be represented as a two dimensional Jones
vector given by
\begin{equation}
|\mathbf{\psi(\theta,\Phi)}\rangle=\cos\frac{\theta}{2}|\mathbf{N}_l\rangle
+\sin\frac{\theta}{2}|\mathbf{S}_m\rangle
\exp(+i\sigma\Phi)\label{HOPS}.
\end{equation}
where
\begin{equation}
|\mathbf{N}_l\rangle=\frac{\sqrt2}{2}(\mathbf{\hat{e}}_x+i\sigma\mathbf{\hat{e}}_y)\exp(il\Phi/2)\label{NL},
\end{equation}
\begin{equation}
|\mathbf{S}_m\rangle=\frac{\sqrt2}{2}(\mathbf{\hat{e}}_x-i\sigma\mathbf{\hat{e}}_y)\exp(im\Phi/2)\label{SM}.
\end{equation}
Here, $(\theta,\Phi)$ are the latitude and longitude on the sphere.
We have introduced the relation $\theta=\delta$ and
$\Phi=2\alpha_0\pm\pi/2$, where the choice of signs depends on the
circular polarization handedness of the input wave.
Equations~(\ref{NL}) and (\ref{SM}) represent orthogonal circular
polarizations with different topological charges $l$ and $m$.

Similar to the previously proposed Poincar\'{e} spheres, the Berry
connection can be written as~\cite{Berry1984}
\begin{equation}
\mathbf{A}=i\langle\psi(\mathbf{R})|\nabla_R|\psi(\mathbf{R})\rangle\label{BerryC}.
\end{equation}
The components of Berry connection can be obtained in
Appendix~\ref{AppB} as
\begin{equation}
\mathbf{A}_\rho=0\label{BerryCRho},
\end{equation}
\begin{equation}
\mathbf{A}_\theta=0\label{BerryCTheta},
\end{equation}
\begin{equation}
\mathbf{A}_\Phi=-\frac{1}{4\rho\sin\theta}[l(1+\cos\theta)
+(m+2\sigma)(1-\cos\theta)]\label{BerryCPhi}.
\end{equation}
The Berry curvature for the hybrid-order Poincar\'{e} sphere plays
the role of ``magnetic field" in the parameter space and is given by
\begin{equation}
\mathbf{V(R)}=-\nabla_\mathbf{R}\times\mathbf{A}\label{BerryCI}.
\end{equation}
Substituting Eqs.~(\ref{BerryCRho})-(\ref{BerryCPhi}) into
Eq.~(\ref{BerryCI}) we get
\begin{equation}
\mathbf{V(R)}=\frac{l-(m+2\sigma)}{4\rho^2}\mathbf{\hat{\rho}}\label{VRII}.
\end{equation}
Equation~(\ref{VRII}) shows the Berry curvature is proportional to
the variation of total angular momentum of light, a sum of SAM and
OAM.

In Berry's framework a state $\psi(\mathbf{R})$ undergoes a cyclic
transformation over a circuit $C$ in parameter space $\mathbf{R}$,
and then return to the initial state, an additional phase in
addition to dynamic phase arises which is given by
\begin{equation}
\gamma(C)=-\int\int_{C}d\mathbf{S}\cdot\mathbf{V(R)}\label{GPI},
\end{equation}
where
$d\mathbf{S}=\rho^2\sin{\theta}d{\rho}d{\theta}d{\Phi}\mathbf{\hat{\rho}}$~\cite{Berry1987}.
Substituting Eq.~(\ref{VRII}) into Eq.~(\ref{GPI}), the resulting
geometric phase on the hybrid-order Poincar\'{e} sphere is then
given by
\begin{equation}
\gamma(C)=-\frac{l-(m+2\sigma)}{4}\Omega\label{GPII},
\end{equation}
where $\Omega$ is the surface area on the hybrid-order Poincac\'{e}
sphere enclosed by the circuit $C$. Equation (\ref{GPII}) shows that
the geometric phase is directly proportional to the variation of
total angular momenta of light. Interestingly, when the two modes
have the same total angular momentum (as for $q=1$) both the Berry
curvature and the geometrical phase vanish, since the photon
crossing the $q$-plate does not change its total angular
momentum~\cite{Marrucci2008}.

On the plane-wave Poincar\'{e} sphere, the evolution of polarization
states in a homogeneous waveplate can be described as the
transformations of longitude and latitude. A quarter-wave plate can
transform a circular polarization light to a linear polarization
one. This transformation on the plane-wave Poincar\'{e} sphere can
be described as polarization states from the north pole to a point
on the equator, and whose longitude depends on the orientation of
the optical axis. Rotation of the waveplate through an angle
$\alpha$ advances the longitude by an angle $2\alpha$. A half-wave
plate can transform a left-handed circular polarization to a
right-handed one. This transformation is presented by a move from
pole to pole along a great circle. This dependence can be easily
demonstrated if we regard the half-wave plate as two identical
quarter-wave plates~\cite{Padgett1999}.

\begin{figure}
\centerline{\includegraphics[width=8cm]{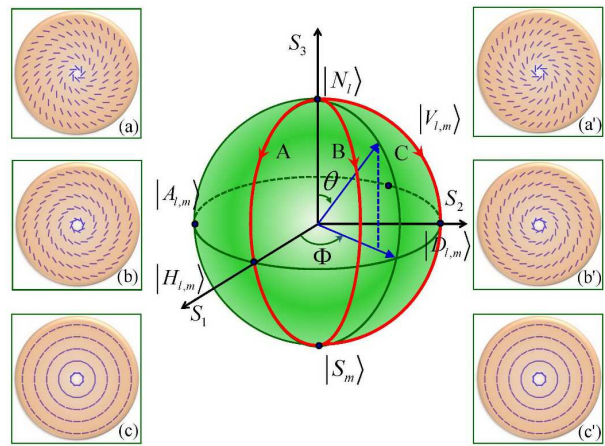}}
\caption{\label{Fig3} (Color online) Schematic illustration of the
evolution of states on the hybrid-order Poincar\'{e} sphere. As an
example, we choose $q=1$ and $l-m=2$. Insets (a)-(c) realization of
the evolution along different longitude lines from the north pole to
south pole by half-wave $q$-plate with different initial angles
$\alpha_0$. Insets (a$'$)-(c$'$) realization of the evolution along
different longitude lines from south pole to north pole. The initial
angle of $q$-plate can be obtained by the relation
$\Phi=2\alpha_0-\pi/2$ from the north pole to south pole, and
$\Phi=2\alpha_0+\pi/2$ from the south pole to north pole.}
\end{figure}

On the hybrid-order Poincar\'{e} sphere, the evolution of
polarization states in an inhomogeneous waveplate can also be
described as the transformations of longitude and latitude. A
quarter-wave $q$-plate transforms a circular polarization light to a
cylindrical vector polarization one. This transformation on the
hybrid-order Poincar\'{e} sphere can be described as polarization
states from the north pole to a point on the equator, and whose
longitude depends on the orientation of the initial angle of
$q$-plate. Rotation of the initial angle $\alpha_0$ of $q$-plate
advances the longitude by an angle $2\alpha_0$ as shown in
Fig.~\ref{Fig3}. Similarly, a half-wave $q$-plate transforms
left-handed circular polarization to right-handed one. This
transformation is presented by a move from one pole to the other
pole along a great circle. Therefore, our scheme provides a
convenient method to describe the spin-orbit interaction. In a word,
the phase retardation of $q$-plate determined the latitude of state
on the hybrid-order Poincar\'{e} sphere, while the longitude is
determined by the initial orientation angle $\alpha_0$. We therefore
can achieve any vector vortex beams by controlling the phase
retardation and initial orientation angle of $q$-plate. An important
point should be noted that for general cases with $q\neq1$,
$\alpha_0$ can be varied by rigidly rotating the $q$-plate, we
therefore can achieve a similar effect of selecting the longitude
line.

For the case of $\sigma=\pm1$ and $l=-m$, the hybrid-order
Poincar\'{e} sphere reduces to the high-order Poincar\'{e}
sphere~\cite{Milione2011,Holleczek2011}. For the case of $\sigma=0$
and $l=-m$, the hybrid-order Poincar\'{e} sphere reduces to the
orbital Poincar\'{e} sphere~\cite{Padgett1999,Galvez2003}.  For the
case of $\sigma=\pm1$ and $l=m=0$ the hybrid-order Poincar\'{e}
sphere reduces to the well-known fundamental plane-wave Poincar\'{e}
sphere. In addition, the hybrid-order Poincar\'{e} sphere can also
be extended to describe the electron vortex beam where the
Pancharatnam-Berry phase is related to real magnetic
field~\cite{Karimi2012}. Note that the rotational symmetry of a
light beam's electric field gives rise to the temporal frequency
shift or rotational Doppler
effect~\cite{Garetz1979,Simon1988,Birula1997,Courtial1998}. For a
spatial rotation of anisotropic axis in plane transverse to the
propagation direction of the beam, the rotational Doppler effect is
valid by replacing the temporal frequency shift with a spatial
one~\cite{Bliokh2008,Niv2008}. The hybrid-order Poincar\'{e} may
provide a convenient route to describe the spatial Doppler effect.

\section{Conclusions}
In conclusions, we have proposed a hybrid-order Poincar\'{e} sphere
to describe the evolution of polarization states in inhomogeneous
anisotropic media. The metasurface (a two-dimensional
electromagnetic nanostructure)~\cite{Kildishev2013} is expected to
be a good candidate for realizing the evolution of polarization
states on hybrid Poinca\'{e} sphere. By correctly controlling the
local orientation and geometrical parameters of the nanograting, one
can achieve any desired polarization distribution on hybrid-order
Poincar\'{e} sphere~\cite{Liu2014,Yi2014}. We have demonstrated that
both the Berry curvature and the Pancharatnam-Berry phase on the
hybrid-order Poincar\'{e} sphere is proportional to the variation of
total angular momentum. A representation of beams in the framework
of the hybrid-order Poincar\'{e} sphere would offer great utility to
describe the spin-orbit interaction and Pancharatnam-Berry phase.

\begin{acknowledgements}
We are sincerely grateful to the anonymous referee, whose comments
have led to a significant improvement of our paper. This research
was partially supported by the National Natural Science Foundation
of China (Grants 11274106 and 11474089) and Foundation of Hubei
Educational Committee (Grant No. Q20132703).
\end{acknowledgements}

\appendix
\section{Calculation of beam evolution in inhomogeneous media} \label{AppA}
In this appendix we give a detailed calculation of beam evolution in
inhomogeneous anisotropic media. The manipulation of polarization
state and phase is obtained by using the effective birefringent
nature of inhomogeneous media. If the orientation of the optical
axis is space-variant at each location, the grating can be described
by the space dependent matrix~\cite{Yariv2007}:
\begin{equation}
\mathbf{T}(r,\varphi)=\mathbf{M}(r,\varphi)\mathbf{J}\mathbf{M}^{-1}(r,\varphi).
\end{equation}
Here, $\mathbf{J}$ is the Jones matrix of a uniaxial crystal, and
\begin{equation}
\mathbf{M}(r,\varphi)=\left(
\begin{array}{cc}
\cos\alpha & \sin\alpha \\
\sin\alpha & -\cos\alpha
\end{array}
\right)\label{Jones},
\end{equation}
where $\alpha(r, \varphi)$ is the local orientation of the optical
axis. It can be easily proved that the Jones matrix
$\mathbf{T}(r,\varphi)$ of the optical field interacting with the
inhomogeneous anisotropic media at each transverse position
$(r,~\varphi)$ is given by following:
\begin{equation}
\mathbf{T}(r,\varphi)=\cos\frac{\delta}{2}\left(
\begin{array}{cc}
1 & 0 \\
0 & 1
\end{array}
\right)-i\sin\frac{\delta}{2}\left(
\begin{array}{cc}
\cos2\alpha & \sin2\alpha \\
\sin2\alpha & -\cos2\alpha
\end{array}
\right)\label{Jones}.
\end{equation}

The Jones vector of electric field associated with input wave is
given by
$|\mathbf{\psi}\rangle={\sqrt2}/2(\mathbf{\hat{e}}_x+i\sigma\mathbf{\hat{e}}_y)
\exp(il\varphi)$. The beam in the inhomogeneous anisotropic media
$|\mathbf{\psi(\delta,\varphi)}\rangle=\mathrm{T}(r,\varphi)|\mathbf{\psi}\rangle$
can be written as
\begin{eqnarray}
|\mathbf{\psi(\delta,\varphi)}\rangle&=&\cos\frac{\delta}{2}\exp(il\varphi)
\frac{\sqrt2}{2}(\mathbf{\hat{e}}_x+i\sigma\mathbf{\hat{e}}_y)-i\sin\frac{\delta}{2}\nonumber\\
&&\times\frac{\sqrt2}{2}(\mathbf{\hat{e}}_x+i\sigma\mathbf{\hat{e}}_y)\exp[i(l\varphi+2\sigma\alpha)]\label{CVVTI}.
\end{eqnarray}
Here, we pay our attention to the evolution of polarization and
phase, and therefore ignoring the evolution of intensity in the
radial coordinate.

\section{Calculation of Berry connection and Berry curvature}\label{AppB}
In this appendix we give a detailed calculation of Berry connection
and Berry curvature. From Eq.~(\ref{BerryC}) the three components of
Berry connection can be written as
\begin{equation}
\mathbf{A}_\rho=i\langle\psi(\mathbf{R})|\partial_\rho|\psi(\mathbf{R})\rangle\label{ARhoI},
\end{equation}
\begin{equation}
\mathbf{A}_\theta=i\langle\psi(\mathbf{R})|\partial_\theta|\psi(\mathbf{R})\rangle/\rho,\label{AThetaI}
\end{equation}
\begin{equation}
\mathbf{A}_\Phi=i\langle\psi(\mathbf{R})|\partial_\varphi|\psi(\mathbf{R})\rangle/(\rho\sin\theta).\label{APhiI}
\end{equation}

As $|\psi(\mathbf{R})\rangle$ is independent of $\rho$, and
$\partial_\rho|\psi(\mathbf{R})\rangle=0$. Substituting it into
Eq.~(\ref{ARhoI}) we get
\begin{equation}
\mathbf{A}_\rho=i\langle\psi(\mathbf{R})|\partial_\rho|\psi(\mathbf{R})\rangle=0\label{ARhoII}.
\end{equation}
From Eq.~(\ref{HOPS}), we know that
\begin{equation}
|\partial_\theta\mathbf{\psi(\mathbf{R})}\rangle
=-\frac{1}{2}\sin\frac{\theta}{2}|\mathbf{N}_l\rangle+\frac{1}{2}\cos\frac{\theta}{2}|\mathbf{S}_m\rangle
\exp(+i\sigma\Phi)\label{FFFI}.
\end{equation}
Substituting it into Eq.~(\ref{AThetaI}) we get
\begin{eqnarray}
\mathbf{A}_\theta&=&i\langle\psi(\mathbf{R})|\partial_\theta|\psi(\mathbf{R})\rangle/\rho\nonumber\\
&=&-\frac{1}{4\rho}\sin\theta\langle\mathbf{N}_l|\mathbf{N}_l\rangle
+\frac{1}{4\rho}\sin\theta\langle\mathbf{S}_m|\mathbf{S}_m\rangle\nonumber\\
&=&0\label{CVVTI}.
\end{eqnarray}
From Eq.~(\ref{HOPS}), we get
\begin{equation}
|\partial_\Phi\mathbf{\psi(\mathbf{R})}\rangle
=\frac{il}{2}\sin\frac{\theta}{2}|\mathbf{N}_l\rangle+\frac{i(m+2\sigma)}{2}\cos\frac{\theta}{2}|\mathbf{S}_m\rangle
e^{+i\sigma\Phi}.\label{FFFII}
\end{equation}
Substituting it into Eq.~(\ref{APhiI}) we get
\begin{eqnarray}
\mathbf{A}_\Phi&=&i\langle\psi(\mathbf{R})|\partial_\Phi|\psi(\mathbf{R})\rangle/(\rho\sin\theta)\nonumber\\
&=&-\frac{1}{4\rho\sin\theta}[l(1+\cos\theta)
+(m+2\sigma)(1-\cos\theta)].\nonumber\\\label{APhiII}
\end{eqnarray}

The Berry curvature is given by
$\mathbf{V(R)}=-\nabla_\mathbf{R}\times\mathbf{A}$, where the
Laplace operator in the sphere coordinate representations can be
written as
\begin{equation}
\nabla=\frac{d}{d\rho}\mathbf{\rho}+\frac{1}{\rho}\frac{d}{d\theta}\mathbf{\theta}
+\frac{1}{\rho\sin\theta}\frac{d}{d\Phi}\mathbf{\Phi}.\label{Laplace}
\end{equation}
We then get
\begin{equation}
\mathbf{V(R)}=\left|
\begin{array}{ccc}
\mathbf{\rho} & \rho\mathbf{\theta} & \rho\sin\theta\mathbf{\Phi}\\
\frac{d}{d\rho} & \frac{d}{d\theta}& \frac{d}{d\Phi}\\
\mathbf{A}_\rho & \rho\mathbf{A}_\theta&
\rho\sin\theta\mathbf{A}_\Phi
\end{array}
\right|\label{LaplaceM}.
\end{equation}
Because of $\mathbf{A_\rho}=\mathbf{A_\theta}=0$, we can just obtain
a component from Eq.~(\ref{LaplaceM}) as
\begin{equation}
\mathbf{V_\rho}=\frac{\frac{d}{d\rho}(\rho\sin\theta\mathbf{A}_\Phi)}{\rho^2\sin\theta}.\label{VRho}
\end{equation}
After substituting Eq.~(\ref{APhiII}) into Eq.~(\ref{VRho}), we get
\begin{equation}
\mathbf{V_\rho(R)}=\frac{l-(m+2\sigma)}{4\rho^2},
\end{equation}
which is proportional to the variation of total angular momenta of
light.

\end{document}